\documentclass[prd,aps,eqsecnum,floatfix,nofootinbib,preprint,tightenlines]{revtex4}

\usepackage{latexsym}
\usepackage{graphicx}
\usepackage{amsmath}

\def\bibi{\bibitem}

\def\c{\chi}
\def\d{\delta}

\def\m{\mu}
\def\n{\nu}
\def\o{\omega}
\def\p{\pi}                     
\def\r{\rho}                    

\def\L{\Lambda}

\def\S{\Sigma}

\def\cd{{\cal D}}

\def\cl{{\cal L}}

\def\cbo{{\,\raise-.15ex\Sc [\,}}                       

\def\lvec#1{\raisebox{0.0ex}{$\stackrel{\leftarrow}{#1}$} }  

\def\ddt#1{{\buildrel {\hbox{\LARGE .\kern-2pt.}} \over {#1}}}

\def\ie{\mbox{\it i.e.}}

\def\tr{{\rm tr}\,}

\def\ie{{\it i.e.}}

\def\tr{{\rm tr}}

\def\chibar{{\overline{\chi}}}

\def\st{{$\mbox{$s_t$ }$}}
\def\pt{{$\mbox{$p_t$ }$}}
\newcommand{\pref}[1]{(\ref{#1})}
\newcommand{\spr}{s_t}
\newcommand{\ppr}{p_t}
\newcommand{\psibar}{\overline{\psi}}
\newcommand{\tn}{t_0}
\newcommand{\wn}{w_0}
\newcommand{\eet}{\langle E_t \rangle}
\newcommand{\et}{E_t}

\begin{document}
\hyphenation{fer-mio-nic per-tur-ba-tive pa-ra-me-tri-za-tion
pa-ra-me-tri-zed a-nom-al-ous}

\renewcommand{\thefootnote}{$*$}

\hfill HU-EP- 13/79

\hfill SFB/CPP-13-113

\begin{center}
\vspace*{10mm}
{\large\bf Chiral perturbation theory for gradient flow observables}
\\[12mm]
Oliver B\"ar$^a$ and Maarten Golterman$^b$
\\[8mm]
{\small\it
$^a$Department of Physics
\\Humboldt University,
Berlin, Germany}
\\[5mm]
{\small\it
$^b$Department of Physics and Astronomy
\\San Francisco State University,
San Francisco, CA 94132, USA}
\\[10mm]
{ABSTRACT}
\\[2mm]
\end{center}

\begin{quotation}
We construct the next-to-leading-order chiral lagrangian for scalar and pseudo-scalar densities defined using the gradient flow, 
for flow times much smaller than the square of the pion wavelength.   We calculate the chiral condensate
and the pion decay constant to this order from operators at positive
flow time, and confirm results obtained earlier in the chiral limit.   
We also calculate the quark mass dependence of the scales $t_0$
and $w_0$ defined from the scalar gluon density and find that nonanalytic terms in the quark mass only enter at next-to-next-to-leading
order.
\end{quotation}

\renewcommand{\thefootnote}{\arabic{footnote}} \setcounter{footnote}{0}

\newpage
\section{\label{Intro} Introduction}
Recently, the application of gradient-flow techniques to nonperturbative field theory
has been extended to include the quark fields of QCD \cite{Luscher:2013cpa,Luscher:2013vga}.   This makes it possible to use the smoothing
properties of the gradient flow for the computation of simple hadronic quantities,
such as the chiral condensate and the pion decay constant in the chiral
limit from the scalar condensate at positive flow time, which avoids 
mixing with power-divergent lower-dimension operators.

In practice, even if one is only interested in the chiral limit, it is useful
to have access to the functional form for the dependence on the quark mass
of the ratios which are computed in order to obtain such
quantities.   Since the quark mass dependence for low-energy hadronic
quantities is encoded in chiral perturbation theory (ChPT), this makes it
necessary to extend ChPT to include effective operators corresponding
to QCD operators at positive flow time.   Our aim
in this article is to provide a systematic framework for this extension to next-to-leading order (NLO) in the quark mass.

The operators we will consider are the scalar and pseudo-scalar quark
bilinears, as well as the purely gluonic operator $G^a_{\m\n}G^a_{\m\n}$.
The latter is of specific interest, because its dependence on the flow time
can be used to define a scale $t_0$ \cite{Luscher:2010iy} or a scale
$w_0$ \cite{Borsanyi:2012zs} from the expectation value of this quantity.   
Because of the 
statistical accuracy with which these scales can be computed, they are good
candidates for scale setting in QCD \cite{lat13:rainer}.   However, this makes it important 
to also determine their dependence on the quark masses accurately. 
Here we provide the functional form of the quark mass dependence of
the scales $t_0$ and $w_0$ to next-to-next-to-leading order (NNLO),
both in the 2 and 2+1 flavor theories.

The outline of this article is as follows.   In the next section, we review
the necessary elements of the gradient flow as applied to QCD, and
introduce scalar and pseudo-scalar quark bilinears at positive flow time.
In Sec.~\ref{FlowChPT} we construct the chiral lagrangian, including 
source terms for these bilinears, to NLO, and calculate the chiral
condensate and the pion decay constant as proposed in Ref.~\cite{Luscher:2013cpa}.  In Sec.~\ref{flowtime} we analyze the quark mass dependence
of $\langle G^a_{\m\n}G^a_{\m\n}\rangle$ at positive flow time, and
from that the quark mass dependence of $t_0$ and $w_0$, to NNLO.
The final section contains our conclusions.

\section{\label{FlowQCD} The gradient flow in QCD}
We begin by recalling relevant definitions and results from Refs.~\cite{Luscher:2010iy,Luscher:2013vga,Luscher:2013cpa}.  We introduce the observables of interest, and then discuss
their chiral properties, as well as their inclusion in the generating functional.

\subsection{\label{BDef} Gradient flow and observables}
The Yang-Mills gradient flow evolves the gauge fields as a function of an additional parameter $t$, referred to as flow time. The $t$ dependent gauge fields $B_{\mu}(t,x)$ satisfy the first-order differential equation\footnote{Here
we take $B_{\mu}(t,x)$ anti-hermitian, following the convention of Ref.~\cite{Luscher:2013cpa}.}
\begin{eqnarray}
\partial_t B_{\mu} & =& D_{\nu}G_{\mu\nu}\,,\label{DefflowEqa}
\\
G_{\mu\nu}&=& \partial_{\mu}B_{\nu}-\partial_{\nu}B_{\mu} +[B_{\mu},B_{\nu}], \qquad D_{\mu}\,=\,\partial_{\mu} + [B_{\mu}, \,\cdot\,]\,,\label{DefflowEqb}
\end{eqnarray}
with initial conditions relating them to the standard gauge fields, $B_{\mu}(0,x) = A_{\mu}(x)$. The flow time dependent quark and antiquark fields $\chi(t,x)$ and $\overline{\chi}(t,x)$ are defined by 
\begin{equation}
\label{DefflowEqc}
\partial_t \chi = \Delta \chi\,,\qquad \partial_t\overline{\chi} = \overline{\chi} \lvec{\Delta}\,,
\end{equation}
with $\Delta=D_{\mu}D_{\mu}$ the covariant laplacian.  
As for the gauge field the initial conditions 
\begin{equation}
\label{Init}
\chi(0,x)=\psi(x)\,,\qquad \chibar(0,x) = \psibar(x)\,,
\end{equation}
relate the flow time dependent fields to the standard quark and antiquark fields $\psi$ and $\psibar$ at zero flow time.

Flow time dependent composite fields are defined in the same way as with the standard fields. For example, we define the scalar and pseudo-scalar densities 
\begin{equation}
\label{DefDensities1}
S_t^{rs}(x) = \chibar_r(t,x)\chi_s(t,x)\,,\qquad P_t^{rs}(x)= \chibar_r(t,x)\gamma_5 \chi_s(t,x)\,.
\end{equation}
Here the labels $r$ and $s$ are flavor indices. The scalar density is also used to define the flow time dependent quark condensates 
\begin{equation}
\label{Defcondt}
\Sigma^{rr}_t=- \langle S_t^{rr}(x)\rangle\,.
\end{equation}
The field theory with the flow time dependent fields can be formulated as a five dimensional local field theory, the fifth dimension being the flow time \cite{Luscher:2013cpa}. 
In particular, the renormalization properties can be discussed in a transparent way in this higher dimensional setup. One can show the absence of
 short-distance singularities in correlation functions involving the densities at positive flow time other than the usual ones occurring in the renormalization of the QCD
 lagrangian. Moreover,  the quark condensates do not require additive renormalization, and the structure of the Wick contractions in the five dimensional theory together with the PCAC relation lead to the Ward--Takahashi identity (WTI)
\begin{equation}
\label{WI1}
(m_r +m_s) \int{\rm d}^4x\langle P^{rs}(x) P_t^{sr}(y) = -\Sigma^{rr}_t - \Sigma^{ss}_t\,,
\end{equation}
where
$m_{r,s}$ denote renormalized quark masses.\footnote{Throughout we consider renormalized parameters and fields only, so we drop the subscript {\small R} used in Ref.~\cite{Luscher:2013cpa}.}   Using this identity,
Ref.~\cite{Luscher:2013cpa} showed that the chiral condensate $\S$ and the
pion-decay constant $f$ in the chiral limit can be obtained from
\begin{eqnarray}
&&\Sigma = \lim_{m\rightarrow0} \frac{\Sigma^{rr}_t G_{\pi}}{G_{\pi,t}}\,,\label{DefSigma}\\
&& f= \lim_{m\rightarrow0} \frac{\Sigma^{rr}_t}{G_{\pi,t}}\,.\label{DefF}
\end{eqnarray}
Here
$G_{\pi}$ is the standard vacuum-to-pion matrix element of the pseudo scalar density,
\begin{equation}
\label{DefG}
 \langle 0 | P^a(x) | \pi^b(p) \rangle = \delta^{ab}\, i G_{\pi}\, e^{ipx}\,,
\end{equation}
and $G_{\pi,t}$ is the analogous expression with $P^a(x)$ replaced by $P^a_t(x)$.   (For the flavor index structure, see Eq.~(\ref{DefDensities4})
below.)

\subsection{\label{ChProp} Chiral properties}
The quark-flow equations in \pref{DefflowEqc} are trivial in flavor and Dirac space. Hence chiral symmetry is preserved by the evolution in flow time, and the fields at positive $t$ transform the same way under chiral transformations as the ones at $t=0$. The same holds for the composite fields. 
In order to discuss the consequences of this we find it more convenient to define the singlet and nonsinglet densities 
\begin{equation}
\label{DefDensities4}
S_t^{a}(x) = \chibar(t,x)T^a\chi(t,x)\,,\qquad P_t^{a}(x)= \chibar(t,x)\gamma_5T^a \chi(t,x)\,,
\end{equation}
with $T^a, a=1\,\ldots,N^2_f-1$, being the generators of the SU($N_f$) flavor group, normalized according to $\tr(T^aT^b)=\delta^{ab}/2$. With this normalization $T^a=\sigma^a/2$ in the case of $N_f=2$. We also allow for $a=0$ and define $T^0$ to be the identity matrix.

With these definitions we consider an infinitesimal chiral rotation 
for the $\c$ and $\chibar$ fields, valid at all values of $t$ including $t=0$,
\begin{equation}
\label{chiralRot2}
\delta\chi(t,x) = \omega^a T^a\gamma_5\chi(t,x)\,,\qquad \delta\chibar(t,x) = \omega^a\chibar(t,x) \gamma_5T^a\,.
\end{equation}
For the nonsinglet densities we find 
\begin{equation}
\label{TrafoNonSing}
\delta S_t^b =\omega^a \left(\frac{1}{N_f}\,\delta^{ab} P_t^0 + d^{abc} P_t^c\right)\,, \quad \delta P_t^b =\omega^a \left(\frac{1}{N_f}\, \delta^{ab} S_t^0 + d^{abc} S_t^c\right)\,, 
\end{equation}
with $\{T^a,T^b\} =\delta^{ab}/N_f + d^{abc} T^c$, while the singlet densities transform according to 
\begin{equation}
\label{TrafoSing}
\delta S_t^0 =\omega^a 2P_t^a\,, \quad \delta P_t^0 =\omega^a 2  S_t^a\,.
\end{equation}

These transformation properties can be used to derive the chiral Ward identity \pref{WI1} directly in four dimensions. 
To do so we consider the expectation value of ${\cal O}_t$, a product of local composite fields at nonzero flow time. As usual it is given as the functional integral over the gauge and fermion fields,
\begin{equation}
\label{DefEV}
\langle {\cal O}_t \rangle = \frac{1}{Z} \int_{\rm fields} {\cal O}_t \,e^{-S}\,.
\end{equation}
Performing a global chiral transformation with constant $\o_a$ on the integration variables, this leads to the WTI\footnote{
Local transformations lead to an additional term in $\delta S$ containing the divergence of the axial-vector current.  However,
this is not very useful here since the variation $\delta {\cal O}_t$ is not simple. }

\begin{equation}
\label{GenWI}
\langle \delta S\  {\cal O}_t \rangle = \langle \delta {\cal O}_t\rangle\,.
\end{equation}
For ${\cal O}_t = P^b_t(y)$ ($b\ne 0$), and degenerate quark mass
$m$,  Eq.~(\ref{TrafoNonSing}) then leads to
\begin{equation}
\label{WI5}
2m \int {\rm d}^4x \langle P^a(x) P_t^b(y)\rangle  = \frac{1}{N_f}\, \d^{ab}\langle S_t^0(y)\rangle\,,
\end{equation}
which is a reformulation of the WTI~(\ref{WI1}). The additional factor $1/N_f$  appears here because of the difference between the densities~(\ref{DefDensities1}) and~(\ref{DefDensities4}).

The derivation of Eq.~(\ref{WI5}) follows directly from the transformation~(\ref{chiralRot2}); 
no other properties of the gradient flow were used. 
As already mentioned, this WTI was derived in Ref.~\cite{Luscher:2013cpa} in a five dimensional setup. In that derivation more information can be extracted. For example, the absence of any nonintegrable singularities at $x=y$ on the left-hand side can be demonstrated for $t>0$.  This follows from the smoothing property of the flow equations.

\subsection{\label{DefGenF} External fields and the generating functional}
It is straightforward to define
a generating functional for correlation functions involving the densities at nonzero flow time, at some particular value of $t$.  External sources are introduced for all densities  of
interest, and correlation functions are obtained by functional derivatives with respect to these sources. For correlators like the one on the left-hand side of Eq.~(\ref{WI5})  we need to introduce source fields for the densities at 
nonvanishing as well as vanishing flow time \cite{Luscher:2013cpa}. Hence, the lagrangian consists of three parts,
\begin{equation}
\label{DeftotLag}
{\cal L}={\cal L}_{{\rm QCD}, m=0} + {\cal L}_{\rm source} + {\cal L}^{\prime}_{\rm source}\,.
\end{equation}
The first part is the massless QCD lagrangian, while the two source terms read
\begin{equation}
\label{Lsource}
{\cal L}_{\rm source} = \psibar(s + i \gamma_5 p)\psi\,,\qquad {\cal L}^{\prime}_{\rm source} = \chibar(\spr + i \gamma_5 \ppr)\chi\,.
\end{equation}
Here $s$ is a matrix-valued and space-time dependent external field, $s(x) = s^0(x) + s^a(x)T^a$, and  analogous definitions apply to the fields $p$, $\spr$, and $\ppr$.
Functional derivatives of the action $\int {\rm d}^4x\,{\cal L}$ reproduce the previously defined densities. Correlation functions are obtained as functional derivatives of the generating functional
\begin{equation}
\label{DefZQCD}
Z_{\rm QCD}[s,p,\spr,\ppr] = \int {\rm D}[A_{\mu},\psibar,\psi] \, e^{-\int d^4x\,{\cal L}[s,p,\spr,\ppr]}\,.
\end{equation}
We obtain the correlation functions of interest by setting all sources
equal to zero after taking derivatives, except for $s$, which is set equal to
the quark mass matrix.

In the next section we construct the generating functional in the chiral effective theory which reproduces the correlation functions obtained with $Z_{\rm QCD}$ at long distance.  For this construction the symmetry properties of the generating functional are needed. In the absence of the primed source fields $Z_{\rm QCD}$ has the known symmetry properties used in Refs.~\cite{Gasser:1983yg,Gasser:1984gg} to construct standard ChPT. If we add sources for the conserved vector and  axial-vector  currents the generating functional is invariant under local chiral transformations as long as $\spr$ and $\ppr$
vanish. In the presence of these fields, however, the local invariance is reduced to a global invariance. As long as we do not consider the currents this distinction is not relevant, and we just state the transformation behavior of the external fields under global chiral symmetry transformations $L\in {\rm SU}(N_f)_L$ and $R\in {\rm SU}(N_f)_R$:
\begin{subequations}
\begin{eqnarray}
\label{Trafo}
&&s + i p\, \longrightarrow\, L(s + i p)R^{\dagger}\,, \label{Trafoa} \\
&& \spr + i \ppr\, \longrightarrow\, L(\spr + i \ppr)R^{\dagger}\label{Trafob}\,.
\end{eqnarray}
\end{subequations}
Clearly, $s + i p$ and $\spr + i \ppr$ transform the same way under chiral rotations. The same is true under parity and charge conjugation, a simple consequence of the fact that the flow equations respect these symmetries as well.

\section{\label{FlowChPT} Chiral perturbation theory}
The densities at nonzero $t$ are not localized at individual space-time points. Instead, they are smeared over a finite region in space-time, called the {\em footprint}, and this footprint is roughly equal to a sphere with radius $\sqrt{8t}$ \cite{Luscher:2013cpa}.
As long as the footprint is much smaller than the inverse pion mass $M_\p$, \ie 
\begin{equation}
\label{boundfootprint}
8t M^2_{\pi} \ll 1\,,
\end{equation}
the smeared densities are effectively point-like for the pions and can be described by point-like densities in the effective theory. This allows us to construct the generating functional $Z_{\rm eff}$ as a function of the external sources $s$, $p$, $\spr$, and $\ppr$, coupled to local effective fields.

\subsection{\label{EffLag} The effective Lagrangian}
We begin with the construction of the chiral effective lagrangian.
The part involving the point-like fields leads to the well-known ChPT expressions of Ref.~\cite{Gasser:1984gg}. To set our conventions, we repeat briefly the relevant formulae.

To leading order (LO) we have ${\cal L}_{\rm LO} = {\cal L}_2$, where ${\cal L}_2$ contains the terms involving two partial derivatives or one power of the sources:
\begin{equation}
\label{L2point}
{\cal L}_2 = \frac{f^2}{4}\, \tr(\partial_{\mu} U\partial_{\mu} U^{\dagger}) - \frac{f^2}{4}\,\tr(\chi^{\dagger}U + U^{\dagger}\chi)\,,
\end{equation}
with
$U$ the standard nonlinear field 
\begin{equation}
\label{DefU}
U(x)=\exp\left(\frac{2i \pi^a(x) T^a}{f}\right)\ ,
\end{equation}
with $\p^a$ the Goldstone boson (GB) fields, and 
$f$ the pion decay constant in the chiral limit.\footnote{Our convention is 
such that $f_\pi=92.21$~MeV.} The source fields $s$ and $p$ are contained in the quantity $\chi$, defined as usual by\footnote{This $\chi$ is not to be confused with the fermion field of Eq.~(\ref{Init}).  From here on, $\c$
will always refer to the source field used in the construction of the chiral
lagrangian.}
\begin{equation}
\label{Defchi}
\chi=2B(s+ip)\,,
\end{equation}
introducing a second low-energy constant (LEC) $B$. 

The sources \st and \pt have the same transformation properties as the sources $s$ and $p$. 
Therefore, a third term for the LO effective action can be constructed:
\begin{equation}
\label{L2flow}
{\cal L}_2^{\prime}=- \frac{f^2}{4}\tr(\chi^{\prime\dagger}U + U^{\dagger}\chi^{\prime})\,,
\end{equation}
where we introduced, in analogy to \pref{Defchi}, the primed source term
\begin{equation}
\label{Defchip}
\chi^{\prime}=2B^{\prime}(s_t+ip_t)\,.
\end{equation}
It involves a new LEC $B^{\prime}$, which in general differs from $B$ because the coupling of the GBs to the densities at nonzero flow time is in general different from the coupling to the densities at $t=0$. In other words, $B^{\prime}$ is a function of $t$, which only at $t=0$ is equal to $B$.

At NLO we write
\begin{equation}
\label{LNLO}
{\cal L}_{\rm NLO} = {\cal L}_4 +  {\cal L}^{\prime}_4 + {\cal L}^{\prime\prime}_4 \,.
\end{equation}
Here ${\cal L}_4$ denotes the familiar NLO lagrangian,
\begin{eqnarray}
{\cal L}_4 &=& \phantom{+}L_4 \,\tr(\partial_{\mu} U\partial_{\mu} U^{\dagger})\tr(\chi^{\dagger}U + U^{\dagger}\chi)\nonumber\\
& & + L_5 \,\tr\big((\partial_{\mu} U\partial_{\mu} U^{\dagger})(\chi^{\dagger}U + U^{\dagger}\chi)\big)\nonumber\\
&&-L_6 \,\tr(\chi^{\dagger}U + U^{\dagger}\chi)\tr(\chi^{\dagger}U + U^{\dagger}\chi)\nonumber\\
&&-L_7 \,\tr(\chi^{\dagger}U - U^{\dagger}\chi)\tr(\chi^{\dagger}U - U^{\dagger}\chi)\nonumber\\
&&-L_8\,\tr((\chi^{\dagger}U + U^{\dagger}\chi)(\chi^{\dagger}U + U^{\dagger}\chi))\nonumber\\
&&- H_2 \,\tr(\chi^{\dagger}\chi)+\dots\label{L4point}
\end{eqnarray}
We have dropped the terms proportional to $L_1,\dots,L_3$, $L_9$, $L_{10}$, and $H_1$ because they do not contribute to the quantities we will calculate in the next section.

Replacing  $\chi$ by $\chi^{\prime}$ twice in the terms quadratic in $\c$ in Eq.~(\ref{L4point}), we get ${\cal L}^{\prime\prime}_4$. The LECs associated with these terms are all new and we label them also by two additional primes,
$L_6^{\prime\prime}$, $L_7^{\prime\prime}$, $L_8^{\prime\prime}$, and
$H_2^{\prime\prime}$.\footnote{For simplicity, we will refer to $H_2$ as a
LEC as well, even though it is a ``high-energy'' constant.}
Replacing only one $\chi$ by $\chi^{\prime}$ we obtain ${\cal L}^{\prime}_4$. For conventional reasons we choose to define the terms bilinear in $\chi$ and $\chi^{\prime}$ with a factor 2:
\begin{eqnarray}
{\cal L}^{\prime}_4 &=& \phantom{+}L^{\prime}_4 \,\tr(\partial_{\mu} U\partial_{\mu} U^{\dagger})\,\tr(\chi^{\prime\dagger}U + U^{\dagger}\chi^{\prime})\nonumber\\
& & + L^{\prime}_5 \,\tr\big((\partial_{\mu} U\partial_{\mu} U^{\dagger})(\chi^{\prime\dagger}U + U^{\dagger}\chi^{\prime})\big)\nonumber\\
&&-2L^{\prime}_6 \,\tr(\chi^{\prime\dagger}U + U^{\dagger}\chi^{\prime})\tr(\chi^{\dagger}U + U^{\dagger}\chi)\nonumber\\
&&-2L^{\prime}_7 \,\tr(\chi^{\prime\dagger}U - U^{\dagger}\chi^{\prime})\tr(\chi^{\dagger}U - U^{\dagger}\chi)\nonumber\\
&&-2L^{\prime}_8\,\tr((\chi^{\prime\dagger}U + U^{\dagger}\chi^{\prime})(\chi^{\dagger}U + U^{\dagger}\chi))\nonumber\\
&&- H^{\prime}_2 \Big( \tr(\chi^{\prime\dagger}\chi)+\tr(\chi^{\dagger}\chi^{\prime})\Big)\,.\label{L4prime}
\end{eqnarray}
The reason for this factor 2 is trivial. Taking functional derivatives with respect to the sources the terms quadratic in the sources pick up an extra factor 2, which from ${\cal L}^{\prime}_4$ is only reproduced if we include this factor as in Eq.~(\ref{L4prime}).

There are many more terms one can write down, but they are not independent of the ones included in Eq.~(\ref{L4prime}). For example, the term 
$\tr((\chi^{\prime\dagger}U + U^{\dagger}\chi)(\chi^{\dagger}U + U^{\dagger}\chi^{\prime}))$ is a linear combination of the $L^{\prime}_8$, $H^{\prime}_2$
and $H^{\prime\prime}_2$ terms, so there is no reason to add this term. Similarly, double-trace terms like $\tr(\chi^{\prime\dagger}U + U^{\dagger}\chi)\,\tr(\chi^{\dagger}U + U^{\dagger}\chi^{\prime})$ can be expressed as
a linear combination of the double-trace terms already present in ${\cal L}^{\prime}_4$ and ${\cal L}^{\prime\prime}_4$. 

The generating functional in the effective theory is now defined by
\begin{eqnarray}
\label{DefZ}
Z_{\rm eff}[s,p,\spr,\ppr] &=& \int \cd[\pi] \,e^{-S_{\rm eff}[s,p,\spr,\ppr]}\,,\\
S_{\rm eff} &=& \int d^4x\left(\cl_2+\cl_2^\prime+\cl_4+\cl_4^\prime
+\cl_4^{\prime\prime}+\dots\right)\,,\nonumber
\end{eqnarray}
and correlation functions are obtained in the same way as in the underlying theory. 

We end this section with a number of observations.
First, the effective action contains terms with both primed and unprimed source fields. Taking derivatives with respect to the unprimed sources $\c$ and $\c^\dagger$ and setting $\c^\prime$ and $\c^{\prime\dagger}$ equal to zero
(in either order),
one obtains only correlation functions  of the standard $t=0$ densities.   This, of course, is as expected. In contrast, the expressions for the $t$ dependent densities contain a remnant of the unprimed source fields, because the scalar density is set equal to the physical mass matrix. 

A second comment relates to the WTI~(\ref{WI5}). By construction, this WTI is respected in the effective theory, because
we introduced a single spurion field $\chi^{\prime} = 2B^{\prime} (s_t+ip_t)$. Both external fields, $s_t$ and $p_t$ have the same transformation properties, and in principle one could introduce a separate spurion field for each of them. However, in the transition to ChPT the symmetry properties of the densities~(\ref{TrafoNonSing}) and~(\ref{TrafoSing}) will not be satisfied unless the terms with these separate spurion fields have the same LECs. 
This is automatically achieved with the single spurion field $\chi^{\prime} = 2B^{\prime} (s_t+ip_t)$.

Third, a comment on the difference between the primed and unprimed LECs.
For $t=0$ they are of course equal, for example, $B^\prime(t=0)=B$.
For very small $t$, one expects that one may expand (up to logarithmic corrections)
\begin{equation}
\label{smalltexp}
B^\prime=B+(t\L^2_{\rm QCD})B_1+(t\L^2_{\rm QCD})^2B_2+\dots
\end{equation}
For instance, in the context of lattice QCD, one could choose $t\sim a^2$,
and Eq.~(\ref{smalltexp}) would give the leading dependence of $B$ on the
lattice spacing $a$.  In this case, the gradient flow provides a smearing of the
operators in Eq.~(\ref{DefDensities1}) by an amount of order the lattice spacing.    However, the idea of
introducing the gradient flow is to smooth operators by choosing the
footprint to be some physical scale, holding $t$ fixed in the continuum
limit $a\to 0$.   If $t\L^2_{\rm QCD}$ is not small, the expansion~(\ref{smalltexp}) breaks down, and there is no simple relation between
$B^\prime$ and $B$.

Finally, one might also consider the introduction of vector and axial-vector
currents at positive $t$, and introduce sources $v_{t\m}$ and $a_{t\m}$ for these operators, respectively.   However, for $t>0$, these currents are not conserved,
and thus of limited interest.   In particular, many more LECs would appear
in the theory, because   $Z_{\rm eff}$ would not be invariant under
local gauge transformations on $v_{t\m}$ and $a_{t\m}$.   With
$r_{\m}^\prime=v_{t\m}+a_{t\m}$ and $\ell_{\m}^\prime=v_{t\m}-a_{t\m}$, already at LO new
LECs would appear with the combinations $i\,\tr(r_{\m}^\prime U^\dagger\partial_\m U-\ell_{\m}^\prime\partial_\m UU^\dagger)$, 
$\tr(r_{\m}^\prime r_{\m}^\prime+\ell_{\m}^\prime\ell_{\m}^\prime)$
and $\tr(r_{\m}^\prime U^\dagger\ell_{\m}^\prime U)$ which are unrelated to 
the LEC $f^2$ multiplying $\tr(\partial_\m U\partial_\m U^\dagger)$.

\subsection{\label{condensate} The condensate and decay constant in the chiral limit}
A key result of Ref.~\cite{Luscher:2013cpa} is Eq.~(\ref{DefSigma})
relating the chiral condensate in the chiral limit to the $t$ dependent condensate and a ratio of two vacuum-to-pion matrix elements of pseudo-scalar densities. 
With the set up of the previous section the calculation of these quantities in ChPT is straightforward and Eq.~(\ref{DefSigma}) can be checked explicitly. Working to NLO we will also obtain the leading mass dependence of the ratio on the right-hand side
of Eq.~(\ref{DefSigma}), thus making transparent how the chiral limit is approached.

Here we restrict ourselves to two light flavors with a degenerate quark mass $m$,
in accordance with Ref.~\cite{Luscher:2013cpa} in deriving Eq.~(\ref{DefSigma}) and Eq.~(\ref{DefF}).

The chiral condensate $\Sigma_{t,{\rm NLO}}  = - \langle S_t^0 \rangle/N_f$ to NLO is given by 
\begin{eqnarray}\label{SigmatNLO}
\Sigma_{t,{\rm NLO}} &=& f^2 B^{\prime} \left(1 - \frac{3M^2_{\pi}}{32\pi^2f^2} \log\left(\frac{M^2_{\pi}}{\mu^2}\right) + \frac{4M^2_{\pi}}{f^2}\,(4L^{\prime}_{68} + H^{\prime}_2)\right)\,.
\end{eqnarray}
The coefficient $L^{\prime}_{68}$ denotes the combination $2L^{\prime}_6+L^{\prime}_8$ and $M^2_{\pi}=2Bm$ is the LO pion mass.
Omitting the primes from the various coefficients in Eq.~(\ref{SigmatNLO}) yields the NLO result for the standard condensate $\Sigma_{{\rm NLO}} $ at $t=0$. 

The 1-loop results for $G_{\pi}$ and $G_{\pi,t}$  read \begin{subequations}
\label{ResGpi}
\begin{eqnarray}
G_{\pi} &=& fB\left(1 - \frac{M^2_{\pi}}{32\pi^2f^2}\log \left(\frac{M^2_{\pi}}{\mu^2}\right) +\frac{4M^2_{\pi}}{f^2}\,( 4L_{68} -L_{45})\right)\,,\label{ResGpia}\\
G_{\pi,t} &=& fB^{\prime}\left(1 - \frac{M^2_{\pi}}{32\pi^2f^2}\log \left(\frac{M^2_{\pi}}{\mu^2}\right) +\frac{4M^2_{\pi}}{f^2}\,( 4L^{\prime}_{68} -L_{45})\right)\,.\label{ResGpib}
\end{eqnarray}
\end{subequations}
In analogy to $L_{68}^\prime$ we defined the combinations $L_{68} = 2L_6+L_8$, and $L_{45} = 2L_4+L_5$.
Note that the same term proportional to the unprimed LEC combination  $L_{45}$ appears in these results. This correction stems from the wave-function renormalization and is the same for both $G_{\pi}$ and $G_{\pi,t}$.
Therefore, this correction as well as the chiral logarithm cancel in the ratio  $G_{\pi}/G_{\pi,t}$ and we obtain
\begin{equation}
\label{Res1}
\Sigma_{t} \,\frac{G_{\pi}}{G_{\pi,t}}= f^2 B\left(1 - \frac{3M^2_{\pi}}{32\pi^2f^2} \ln\left(\frac{M^2_{\pi}}{\mu^2}\right) + \frac{4M^2_{\pi}}{f^2} (4L_{68} + H^{\prime}_2)\right)\,.
\end{equation}
As expected, the right-hand side reproduces the chiral condensate in the chiral limit, $\Sigma=f^2B$. 
Note that the right-hand side is $t$ dependent as suggested by the left-hand side, but this $t$ dependence stems solely from the coefficient $H^{\prime}_2$.  For instance, in the free-quark theory, it is easy to see
that $B^2H_2^\prime\sim 1/t$ (if the quark mass $m$ is much smaller
than $t^{-1/2}$), while $B^2H_2\sim \L^2$, with $\L$
the UV cutoff.

Let us rewrite Eq.~(\ref{Res1}) in two different ways. First, by introducing a single dimensionful constant $\Lambda_t$ we can write
\begin{equation}
\label{Res2}
\Sigma_{t}\, \frac{G_{\pi}}{G_{\pi,t}}= \Sigma_{\rm LO} \left(1 - \frac{3M^2_{\pi}}{32\pi^2f^2} \ln\left(\frac{M^2_{\pi}}{\Lambda_t^2}\right)\right)\,,
\end{equation}
a result that can already be found in Ref.\ \cite{Luscher:2013vga}. $\Lambda_t$ inherits its $t$ dependence from $H_2^{\prime}$ but depends on the combination $L_{68}$ of unprimed LECs as well. Alternatively we may write
\begin{equation}
\label{Res3}
\Sigma_{t}\, \frac{G_{\pi}}{G_{\pi,t}}= \Sigma_{\rm NLO} \left(1 -  \frac{4M^2_{\pi}}{f^2} \left( H_2 - H^{\prime}_2\right)\right)\,.
\end{equation}
Here we see that the left-hand side reproduces the condensate to NLO times a correction factor that only involves the difference of the two contributing ``high-energy'' constants. 

The second equation, Eq.~(\ref{DefF}), relates the ratio $\Sigma_{t}/{G_{\pi,t}}$  to the decay constant in the chiral limit. 
For this ratio we obtain to NLO
\begin{equation}
\label{Res4}
\frac{\Sigma_{t}}{G_{\pi,t}} = f\left(1- \frac{M^2_{\pi}}{16\pi^2f^2}\log \left(\frac{M^2_{\pi}}{\mu^2}\right) +\frac{4M^2_{\pi}}{f^2}\,( L_{45} + H_2^{\prime})\right)\,,
\end{equation}
and the expected chiral limit is reproduced. As for the chiral condensate the only $t$ dependence of the ratio comes from $H_2^{\prime}$.
Using the known one-loop result $f_{\pi,{\rm NLO}}$ \cite{Gasser:1983yg} we can rewrite Eq.~(\ref{Res4}) as
\begin{equation}
\label{Res5}
\frac{\Sigma_{t}}{G_{\pi,t}} = f_{\pi,{\rm NLO}} \left(1- \frac{4M^2_{\pi}}{f^2} \,H^{\prime}_2 \right)\,.
\end{equation}
In summary, our results confirm Eqs.~(\ref{DefSigma}) and~(\ref{DefF}). A perhaps unexpected observation is that only one new coefficient $H_2^{\prime}$ enters the ratio $\Sigma_{t}/{G_{\pi,t}}$ in Eqs.~(\ref{DefSigma}) and~(\ref{DefF}), even though many more are present in the chiral lagrangian, {\it c.f.} Eq.~(\ref{L4prime}).  

\section{\label{flowtime} Chiral expansion for the scales $t_0$ and $w_0$}
An important application of the gradient flow is the definition of scales used for scale setting in lattice QCD. Two closely related scales have recently been put forward, $\tn$ \cite{Luscher:2010iy} and $\wn$ \cite{Borsanyi:2012zs}. The starting point in both cases is $\langle E_t \rangle$, the expectation value of the scalar gauge-field density at nonzero flow time,
\begin{equation}
\label{Et}
E_t(x)= \frac{1}{4}\, G^a_{\mu\nu}(t,x) G^a_{\mu\nu}(t,x)\,.
\end{equation}
In terms of $\eet$ the scales $\tn$ and $\wn$ are then defined implicitly by the equations
\begin{equation}
\label{Defscales}
t^2 \eet{\Big|}_{t=t_0}=0.3\,,\qquad
t\frac{d}{d{t}} \{t^2 \eet\}{\Big|}_{t=w_0}=0.3\,.
\end{equation}
These two scales share a few advantages that make them prime candidates for scale setting: 
They are numerically cheap to compute and can be computed very accurately, with a statistical error at the per-mille level. In addition, they show a weak dependence on the quark masses, and therefore quark mass uncertainties have a small effect on these scales. 

In the following we want to show another advantage of these scales: Unlike some other scales, such as the Sommer parameter $r_0$ \cite{Sommer:1993ce}, the flow-dependent scales allow us to calculate their dependence on the quark masses in ChPT. This is because $\et$ is a local
(at the scale of the pion), gauge-invariant observable that can be mapped onto the effective theory just like the densities discussed in the previous section. 

First of all, a source term $\int_x \rho(x)\et(x)$ can be added to the QCD action in the familiar fashion. The expectation value $\eet$ is then obtained by a single functional derivative with respect to the source $\rho(x)$. Since $\et$ is a gluonic observable, it is invariant under chiral transformations, in addition to being a scalar under rotations and parity. This implies that the source $\rho$ is a scalar too. Therefore, in order to represent $\et$ in ChPT we simply have to write down the most general scalar in terms of the chiral field $U$ and the sources $\chi$ and $\c^\prime$. Of course, the constraint that the footprint of $\et$ is much smaller than the Compton wavelength of the pion needs to be satisfied here as well.

This task has essentially been done in the previous section when we constructed the chiral lagrangian. However, since an overall additive constant is irrelevant for the effective action we dropped this constant in Sec.~\ref{EffLag}. Restoring it, we are led to the following chiral expansion for $\et$:
\begin{eqnarray}\label{etchpt}
\et &= &c_1f^4 +c_2f^2 \,\tr(\partial_{\mu} U\partial_{\mu} U^{\dagger}) + c_3f^2 \,\tr(\chi^{\dagger}U + U^{\dagger}\chi)\nonumber \\
& & +\, c_4 \,\tr(\chi\chi^{\dagger}) +c_5 \big[\tr(\chi^{\dagger}U + U^{\dagger}\chi)\big]^2 + c_6 \,\tr(\chi^{\dagger}U\chi^{\dagger}U + U^{\dagger}\chi U^{\dagger}\chi)\nonumber \\
&&+\,c_7\big[\tr(\chi^{\dagger}U - U^{\dagger}\chi)\big]^2\,.
\end{eqnarray}
Since $\et$ is of mass dimension four we make this dimension explicit by inserting appropriate powers of the decay constant in the chiral limit. Hence, the LECs $c_i$ in Eq.~(\ref{etchpt}) are all dimensionless.\footnote{Strictly speaking, the constants $c_1$ and $c_4$ are high-energy constants. However, for brevity  we will refer to all $c_i$ as LECs in this section.} All terms in Eq.~(\ref{etchpt}) except for the one proportional to $c_1$ have an analogue in the chiral lagrangian: the $c_2$ and $c_3$ terms correspond to the two terms in ${\cal L}_2$, while the remaining three correspond to the last four terms in ${\cal L}_4$, {\it c.f.} Eq.~(\ref{L4point}).\footnote{Except for a slight redefinition of the $L_8$ and $H_2$ terms.} Note that we have not transcribed the terms proportional to $L_1,\dots, L_5$, since these will not contribute to the order we are working here.   Also, we have already set $\c^\prime=0$, since no
derivative with respect to $\c^\prime$ is needed.

It is straightforward to expand $\et$ in powers of the pion fields to quadratic order. To be specific let us consider first the case of $N_f=2$ with a degenerate quark  mass $m$. In terms of the tree-level pion mass $M^2_{0} = 2Bm$ we find
\begin{eqnarray}
\label{etnf2}
\et&=& c_1f^4\left( 1+ 4\tilde{c}_3\,\frac{M^2_{0}}{f^2} + 2\tilde{c}_2\, \frac{1}{f^4} \,\partial_{\mu}\pi^a\partial_{\mu}\pi^a- 2\tilde{c}_3\,\frac{M^2_{0} }{f^4} \,\pi^a\pi^a +2\tilde{c}_{456} \,\frac{M^4_{0}}{f^4}\right).
\end{eqnarray}
The coefficients $\tilde{c}_i$ differ from the $c_i$ by a factor of $c_1$, \ie, $\tilde{c}_i=c_i/c_1$, and the last coefficient $\tilde{c}_{456}=\tilde{c}_4+8\tilde{c}_5+2\tilde{c}_6$.  Already here we can make a few comments concerning the chiral expansion. The LO contribution is given by the constant piece $c_1f^4$. The term proportional to $4\tilde{c}_3$ is the NLO part which is suppressed by one power $M^2_{0}/f^2$. The contributions involving two powers of the pion fields will result in terms proportional to the pion propagator at zero distance, hence these terms produce chiral logarithms. The last term will function as the counterterm for the divergences originating from the pion-propagator contributions. Therefore, the nonanalytic mass dependence starts at NNLO. 

Having made these remarks the NNLO result for $\eet$ reads 
\begin{equation}
\label{eetnf2}
\eet= c_1f^4\left( 1+ 4\tilde{c}_3\,\frac{M^2_{0}}{f^2} - 6\tilde{c}_{23}\,\frac{M^4_{0}}{16\pi^2 f^4} \log\left(\frac{M^2_{0}}{\mu^2}\right) +2\tilde{c}^r_{456} \,\frac{M^4_{0}}{f^4}\right)\,.
\end{equation}
Here the coefficient $\tilde{c}_{23} = \tilde{c}_2 + \tilde{c}_3$. The renormalized coefficient $\tilde{c}^r_{456}$ depends on the renormalization scale $\mu$ and cancels the $\mu$-dependence of the chiral logarithm, as usual.    Just as $H_2^\prime$ falls like an inverse power of $t$, $c_1\sim 1/t^2$, at least in perturbation theory \cite{Luscher:2010iy}.

The result \pref{eetnf2}  can now be used in the definitions for $\tn$ and $\wn$.  
For vanishing quark mass we find the scale $t_{0,{\rm ch}}$ in the chiral limit from \begin{equation}
\label{tnnf2cl}
t_{0,{\rm ch}} =\frac{1}{f^2}\sqrt{\frac{3}{10 c_1(t_{0,{\rm ch}})}}\,.
\end{equation}
For a nonvanishing quark mass we expand around $t_{0,{\rm ch}}$ and find the expression
\begin{equation}
\label{tnnf2}
t_0= t_{0,{\rm ch}} \left( 1+ \tilde{k}_1\,\frac{M^2_{0}}{f^2} + \tilde{k}_{2}\,\frac{M^4_{0}}{16\pi^2 f^4} \log\left(\frac{M^2_{0}}{\mu^2}\right) + \tilde{k}_{3} \,\frac{M^4_{0}}{f^4}\right).
\end{equation}
The coefficients $\tilde{k}_j$ are combinations of the coefficients in Eq.~\pref{eetnf2} and their derivatives with respect to $t$. The latter appear because the coefficients $\tilde{c}_j$ are $t$-dependent.  Hence, by defining the scale $\tn$ via the implicit Eq.~\pref{Defscales} the coefficients $\tilde{c}_j=\tilde{c}_j(\tn)$ become mass dependent because $\tn$ is mass dependent. In order to make all mass-dependence explicit we need to expand the coefficients around their values $\tilde{c}_j(t_{0,{\rm ch}})$ in the chiral limit.  
The chiral expansion for $w_0$ has the same form as in Eq.~\pref{tnnf2} but with different coefficients.

The result~(\ref{tnnf2}) is expressed in terms of the tree-level pion mass $M^2_0$. In order to replace $M_0^2$ with $M_{\pi}^2$ we need the NLO result for the pion mass in the NLO correction in $t_0$, while the LO result is sufficient for the NNLO correction. Doing the replacement $2\tilde{c}_3M_0^2/f^2 \rightarrow 2\tilde{c}_3M_{\pi}^2/f^2 $ additional NNLO corrections are spawned. The form of these extra terms is the same as for the already existing NNLO terms  and they can be combined. The only difference with Eq.~(\ref{tnnf2}) is that the NNLO LECs change.  Therefore, in terms of the physical pion mass the final result can be written as
\begin{equation}
\label{tnn2f2}
t_0= t_{0,{\rm ch}} \left( 1+k_1 \,\frac{M^2_{\pi}}{(4\pi f)^2} + k_2 \,\frac{M^4_{\pi}}{(4\pi f)^4} \log\left(\frac{M^2_{\pi}}{\mu^2}\right) + k_3 \,\frac{M^4_{\pi}}{(4\pi f)^4}\right)\,,
\end{equation}
with $k_{1,2,3}$ unknown constants independent of the quark mass.
In writing this expression, we have  inserted factors $1/16\pi^2$ such that the dimensionless expansion parameter appears as $M^2_{\pi}/16\pi^2 f^2$, as usual.
With $k_3 = k_2\, \log\left({\mu^2}/{\Lambda^2_{t_0}}\right)$
this can also be written as 
\begin{equation}
\label{tnn3f2}
 t_0= t_{0,{\rm ch}} \left( 1+k_1 \,\frac{M^2_{\pi}}{(4\pi f)^2} + k_2 \,\frac{M^4_{\pi}}{(4\pi f)^4} \log\left(\frac{M^2_{\pi}}{\Lambda_{t_0}^2}\right)\right)\,.
\end{equation}
As already mentioned, the result for $w_0$ is of the same form as Eq.~(\ref{tnn2f2}) but each term has a different coefficient. 

Let us repeat the crucial feature of our result: the nonanalytic chiral logarithms enter first at NNLO, and thus they are expected to be strongly suppressed. This expectation is supported by data from the Alpha Collaboration \cite{Bruno:2013gha}. Figure~4 of this reference shows the pion mass dependence of $t_0$, and this dependence is linear to a very good approximation. A rough estimate from the data gives $k_1 \approx -1$.

It is straightforward to repeat the calculation for the $N_f=2+1$ case.  
In this case, the requirement analogous to Eq. (3.1) is that $8M_\eta^2 t\ll 1$.  We find 
\begin{eqnarray}
\label{tnn2f3}
t_0&=& t_{0,{\rm ch}} \left( 1+k_1 \,\frac{2M_K^2 + M^2_{\pi}}{(4\pi f)^2}\right. \nonumber\\
&& +  \left.\frac{1}{ (4\pi f)^2} \left((3k_2-k_1) M^2_{\pi}\mu_{\pi} + 4 k_2 M^2_{K}\mu_{K} + \frac{k_1}{3}\left(M_{\pi}^2-4M_K^2\right)\mu_{\eta} +  k_2 M^2_{\eta} \mu_{\eta} \right)\right.\nonumber\\
&&\left. +k_4\,\frac{(2M_K^2+M_{\pi}^2)^2}{(4\pi f)^4} + k_5 \,\frac{(M_K^2-M_{\pi}^2)^2}{(4\pi f)^4}\right)\,,
\end{eqnarray}
with the short-hand notation 
\begin{equation}
\label{logsshort}
\mu_P=\frac{M_P^2}{(4\p f)^2}\log\left(\frac{M^2_P}{\mu^2}\right)\,,\qquad
P=\p,\ K,\ \eta\ .
\end{equation}  
As expected, the NNLO mass dependence is more complicated in the $2+1$ flavor case. 
The leading mass dependence at NLO is analytic and has a very simple form; it is proportional to the combination $2M^2_{K} + M^2_{\pi}$. This result is easy to understand. It stems from the term proportional to $c_3$ in Eq.~(\ref{etchpt}) with $U=U^{\dagger}=1$. At this order, 
$\tr(\chi^{\dagger}U + U^{\dagger}\chi)$  is thus proportional to the sum of the quark masses, which at LO is proportional to $ 2M^2_{K} + M^2_{\pi}$.

Note that there is some arbitrariness in parametrizing the analytic mass dependence at NNLO. The combination we have used has the advantage that the $k_5$ contribution vanishes in the SU(3) flavor-symmetric point $m=m_s$. 

Corrections due to a finite space-time volume \cite{Gasser:1987zq} can be easily included in our results. The finite volume (FV) corrections essentially amount to a simple replacement of the chiral logarithms, $\mu_P \,\rightarrow \,\mu_P + \delta_{\rm FV}$ (see for instance Refs.~\cite{Golterman:2009kw,Colangelo:2005gd}).
Since the chiral logarithms enter at NNLO the FV corrections start at this order too and are expected to be very small. At least for the scale $w_0$ it has been observed that the FV corrections are tiny \cite{Borsanyi:2012zs}.

\section{\label{conclusion} Conclusion}
In this article, we extended chiral perturbation theory to include 
quark bilinears as well as the gluon condensate smoothed using the
gradient flow.   In order for this to be possible,  the square root of the
flow time $t$ has to be much smaller than the Compton wavelength, so that
these operators are local relative to the effective theory \cite{Luscher:2013cpa}.

We calculated the chiral condensate and the pion decay constant
from the ratios given in Eqs.~(\ref{DefSigma}) and~(\ref{DefF}).  As
anticipated in Ref.~\cite{Luscher:2013vga}, we find that a new low-energy
constant appears in the ratio $\S_tG_\p/G_{\p,t}$, {\it c.f.} Eq.~(\ref{Res2}).
The difference between this ratio and the chiral condensate $\S$ at $t=0$ is the difference
between the ``high-energy'' constants $H_2$ and $H_2^\prime$.
We observe that $H_2^\prime\ne 0$, even though it is a finite
quantity.   

We also calculated the quark mass dependence of the gluon
condensate at positive flow time, finding that this dependence is
linear at NLO, with logarithmic terms only entering at NNLO.
This adds to the usefulness of the scales $t_0$ and $w_0$,
because it implies that at low enough quark mass a linear 
extrapolation to the physical point should be quite accurate.
With the estimate $k_1\approx -1$ \cite{Bruno:2013gha},
the NLO term in $t_0$ given in Eq.~(\ref{tnn3f2}) is about 3\% for
a 200~MeV pion.   If we take $\L_{t_0}=M_\r$, the NNLO quantity
$(M_\p/(4\p f))^4\log{M_\p^2/\L_{t_0}^2}$ in Eq.~(\ref{tnn3f2}) varies 
by less than 1\% between a 300~MeV and the physical pion mass,
and less than 0.2\% between a 200~MeV and the physical pion mass. 

We did not extend the effective theory to include correlation
functions defined with vector or axial-vector currents at positive
flow time.   While this extension is straightforward, it is less
interesting, as these currents are not conserved.   As discussed
in Sec.~\ref{EffLag}, this extension would lead to the introduction of 
quite a few more low-energy constants in the effective theory.

\vspace{3ex}
\noindent {\bf Acknowledgments}
\vspace{3ex}

We thank Yigal Shamir and Rainer Sommer for discussions and Claude Bernard for pointing out a mistake in the first version of this paper.   OB is supported in part by the Deutsche Forschungsgemeinschaft (SFB/TR 09), and MG is supported in part by the US Department of Energy.


\end{document}